\documentclass[ 5p, 12pt, twocolumn, times]{elsarticle}
\biboptions{sort&compress}

\usepackage{amsmath}
\usepackage{siunitx}
\usepackage{hyperref}
\makeatletter
\newcommand*{\rom}[1]{\expandafter\@slowromancap\romannumeral #1@}
\makeatother\newcommand*\diff{\mathop{}\!\mathrm{d}}
\usepackage{upgreek}
\usepackage{graphicx}
\usepackage{subfigure}
\usepackage{color}
\usepackage{caption}
\usepackage{tabularx}
\usepackage[T1]{fontenc}

\DeclareMathAlphabet{\Ibb}{U}{msb}{m}{n}
\newcommand   {\IC}{{\ensuremath{\Ibb C}}}
\newcommand \ggp {{\ensuremath{{{\gamma}/{\gamma}^{\prime}}}}}
\newcommand \gp  {{\ensuremath{{\gamma}^{\prime}}}}
\newcommand \g   {{\ensuremath{\gamma}}}
\newcommand{\matone}{\ensuremath{\text{\textup{\textbf{I}}}}}

\newcommand{\Bg}{{\boldsymbol{\mathnormal g}}}

\newcommand{\Bn}{{\boldsymbol{\mathnormal n}}}
\newcommand{\Bu}{{\boldsymbol{\mathnormal u}}}
\newcommand{\Bv}{{\boldsymbol{\mathnormal v}}}

\newcommand{\BJ}{{\boldsymbol{\mathnormal J}}}

\newcommand{\BM}{{\boldsymbol{\mathnormal M}}}

\newcommand{\Bepsilon} {\ensuremath{\boldsymbol\epsilon}}

\newcommand{\allphi}{\phi_1,...,\phi_4}

\renewcommand{\eqref}[1]{Eq.~(\ref{#1})}
\newcommand{\figref}[1]{Fig.~\ref{#1}}
\newcommand{\tabref}[1]{Table~\ref{#1}}

\begin{document}

\begin{frontmatter}


\title{Cyclic-loading microstructure-property relations from a mesoscale perspective: An example of single crystal Nickel-based superalloys}

\author[NWPU,FAU]{Ronghai Wu\corref{corr}}
\corref{mycorrespondingauthor}
\cortext[corr]{Corresponding author.}
\ead{ronghai.wu@nwpu.edu.cn}
\author[FAU,SWJTU]{Michael Zaiser}

\address[NWPU]{School of Mechanics, Civil Engineering and Architecture, Northwestern Polytechnical University, Xian, 710129, PR China}
\address[FAU]{WW8-Materials Simulation, Department of Materials Science, Friedrich-Alexander Universit\"at Erlangen-N\"urnberg, Dr.-Mack-Str. 77, 90762 F\"urth, Germany}
\address[SWJTU]{Southwest Jiaotong University, School of Mechanics and Engineering, Chengdu, PR China}

\begin{abstract}
Past models of stress-strain response under cyclic loading mainly rely on macroscopic equations which consider microstructure evolution indirectly or simply discard microstructure information. Modern materials science, on the other hand, seeks quantitive descriptions for the relations between microstrucutre and loading response. In the present work, we show a promising mesoscale phase-field framework which can describe co-evolution of phase/grain and defect microstructures, reveal microstructure mechanisms and simultaneously predict deformation properties as a natural outcome of microstrucuture interactions. The energy functionals for phase/grain and defect microstructures are constructed, followed by functional variation which leads to governing equations. Applying the developed framework to high temperature cyclic loading of single crystal Nickel-based superalloys, the simulated results show that cyclic loading-microstructure-property relations can be principally revealed. In the short term perspective (in one cycle), dislocations move back and forth, leading to cyclic loops consistent with characteristics observed in experiments. The plastic strains are one order of magnitude smaller than total strains, which explains why the cyclic loops are very "thin". In the long term perspective, all $\ggp$ microstructures exhibit directional coarsening similar to creep under zero cyclic loading ratio, with the extent of rafting slight dependent on cyclic waveform, period, etc. The plastic strains are sensitive to cyclic loading conditions both in terms of curve shape and in terms of magnitude.
\end{abstract}


\end{frontmatter}

\section{Introduction}
The stress-strain responses of metals under cyclic loading have attracted massive investigations in the past decades. Modern experimental techniques can characterize microstructures and test mechanical properties, while modelling and simulation act as bridges between them. Most modelling work on cyclic stress-strain responses focuses on the macroscale which is a scale larger than the characteristic sizes of grain, phase and dislocation structures \citep{Tong_2003_IJF, Tong_2004_IJF, Cornet_2011_IJF, Kang_2006_MSEA, Kang_2012_MSEA, Zhu_2017_IJP}. One of the main advantages is that the complex microstructure interactions are largely simplified. The microstructure information could be incorporated into constitutive models by presumed evolution equations of microstrucutres, with the assistance of available experimental data on corresponding microstructures. Alternatively, one may completely discard microstructure information and develop purely phenomenological constitutive models based on mathematical formulations similar to stress-strain curves. This is reasonable given that computers were invented in the middle of 20st century and the capabilities of computers were weak until the 21st century: simulations below macroscale were not available at that time, and stress-strain curves obtained by macroscopic models can already match experimental data well. Nevertheless, the loading-microstructure and microstructure-property relations cannot be revealed in macroscopic models. In other words, macroscopic models essentially aim at directly reproducing stress-strain curves under different loadings (i.e. loading-property relations) close to experiments, shadowing the mechanisms behind. As the computational power increases, many lower scale methods (e.g. molecular dynamics) are able to provide detailed information on microstructures but are not applicable for collective microstructure evolution or property prediction at engineering levels \citep{Soppa_2014_MSEA, Fan_2016_Acta}. A question may arise: can we find an approach which retains essential underlying physics such that collective microstructure information is meaningfully represented and further microstructure evolution mechanisms can be revealed, and which can simultaneously predict deformation properties as a natural outcome of microstrucuture interactions? Some attempts have been made on creep deformation of metals by the current authors \citep{Wu_2016_SM, Wu_2017_JAC, Wu_2017_IJP}. The present work gives a possible solution -- 
a phase-field approach for cyclic loading on the mesoscale -- to this question.

The present work will first show that both the evolutions of phase/grain and dislocation microstrucures can be cast into a phase-field framework. The capability of the framework will be shown by the example of cyclically loaded single crystal Nickel-based superalloys. 
       
\section{Methods}
\subsection{General phase-field framework for phase/grain and dislocation microstructure evolution }
The essential featuure of phase-field models is that microstructures are represented by continuum quantities whose temporal evolutions link to variation of energy functional of the system. The physical robustness of phase-field evolution equations has been widely accepted: Cahn-Hilliard equation for transport and Allen-Cahn equation for non-transport (e.g. local order/disorder) processes. Feor a system of phases characterized by concentration variables $c_i$ and order parameters $\phi_i$ (characterizing different types of crystallogra[hic order), coupled to a set of dislocation densities, we have
\begin{gather}
\label{eq:c_evolution1}
\partial_t c_i = \nabla \cdot (-\BJ_{c_i}) = \nabla \cdot \Big( M_{c_i} \frac{\delta E^{\rm pha}}{\delta c_i} \Big),\\
\label{eq:phi_evolution1}
\partial_t \phi_i = M_{\phi_i} \frac{\delta E^{\rm pha}}{\delta \phi_i},\\
\label{eq:rho_evolution1}
\partial_t \rho_{i,\pm} = \nabla \cdot (-\BJ_{\rho_{i,\pm}}),
\end{gather}


where $\rho_{i,\pm}$ is a density of positive and negative dislocations in $i$th slip system and $E$ is the energy functional which is shared between the different fields. The evolutions of $c_i$ and $\rho_{i,\pm}$ are essentially the transport equations of zero-dimensional objects
(e.g. atoms and vacancies) and one-dimensional defects (i.g. dislocations), respectively, therefore they follow the Cahn-Hilliard equation. In contrast, $\phi_i$ represents local order/disorder, hence it follows the Allen-Cahn equation. The mobility coefficients (i.e. $M_{c_i}$ and $M_{\phi_i}$) are assumed to be isotropic for simplicity, while their counter-parts for dislocation variables need to reflect the crystallographic anisotropy of dislocation motion. 

The next step is to construct the energy functionals. The structure of the phase-field energy functional of phase/grain microstructures has been well developed; it includes a bulk and a gradient energy term as basic ingredients \citep{Cahn_1958_JCP}. In solids, phase changes are in general associated with inelastic deformations $\Bepsilon^{\rm inel}(c_i,\phi_i)$ which in turn may induce internal stresses, therefore, elastic energy terms need to be incorporated into the energy functional. Thus, the energy functional of phase/grain microstructures can be constructed as:

\begin{equation}
\label{eq:phase_energy_functional}
E^{\rm pha} = \int ({\cal E}^{\rm bulk}(\phi_i,c_i) + {\cal E}^{\rm grad}(\phi_i,c_i,\nabla\phi_i,\nabla c_i) + {\cal E}^{\rm el}) \diff V
\end{equation}
where the specific form of ${\cal E}^{\rm bulk}$ can be directly linked to Calphad or phenomenological polynomials \citep{Zhu_2004_AM} and the elastic energy density has the generic form 
\begin{equation}
\label{eq:elastic_energy_density}
 {\cal E}^{\rm el} = \frac{1}{2} \Bepsilon^{\rm el}:\IC:\Bepsilon^{\rm el}.
\end{equation} 
where $\IC$ is Hooke's tensor of elastic constants. We can express the elastic strain $\Bepsilon$ as a functional of the phase variables. In the limit of small strains this is simply done by setting $\Bepsilon^{\rm el} = \Bepsilon -\Bepsilon^{\rm inel}(c_i,\phi_i)$, where the total strain $\Bepsilon$ is the symmetrized gradient of the displacement field $\Bu$, $\Bepsilon = \nabla \Bu_{\rm sym}$. 

We now turn to the description of the dislocation system and the associated plastic strain. We build upon a continuum dislocation dynamics (CDD) framework developed by systematic averaging of discrete dislocation dynamics (DDD) \citep{Groma_1997_PRB, Zaiser_2001_PRB, Groma_2003_AM}. As shown by \citet{Groma_2007_PM, Zaiser_2015_PRB, Groma_2016_PRB}, this framework can be cast into the form of a phase field model. Here we use a two-dimensional formulation, describing the motion of straight edge dislocations of positive and negative signs in the plane spanned by the slip plane normal $\Bn_i$ and the glide direction $\Bg_i$ and giving rise to a plane strain deformation. This is a serious simplification which, however, it makes the analogy of our dislocation model to the general phase field formalism most evident because this simplification allows us to treat the dislocation densities in strict analogy to conserved, concentration-like variables. 

Starting from \eqref{eq:rho_evolution1}, and accounting for the anisotropy of dislocation motion which occurs in the crystallographic glide  directions $\Bg_i$ only, we obtain 
\begin{gather} 
\label{eq:rho_plus_evolution1}
\BJ_{\rho_{i,+}} =   \BM_{i,+} \nabla \Big[ \frac{\delta E^{\rm dis}}{\delta \rho_{i,+}} \Big],\\
\label{eq:rho_minus_evolution1}
\BJ_{\rho_{i,-}} =   \BM_{i,-} \nabla \Big[ \frac{\delta E^{\rm dis}}{\delta \rho_{i,-}} \Big].
\end{gather}
The anisotropic mobility tensors $\BM_{\pm,i}$ have the form
\begin{equation}
\BM_{i,\pm} = \frac{b}{B({c,\phi})} \rho_{i,\pm} [\Bg_i \otimes \Bg_i].
\end{equation}
where $b$ is the magnitude of Burgers and $B$ is a dislocation drag coefficient which in general depends on the phase composition of the material. With this form of the mobility tensor we can alternatively write the fluxes as
\begin{gather} 
\label{eq:rho_plus_evolution1a}
\BJ_{\rho_{i,\pm}} =  \rho_{i,\pm} \Bv_{i,\pm},\\
\label{eq:rho_minus_evolution1a}
\Bv_{i,\pm} = \Bg_i \frac{b}{B} (\Bg_i.\nabla) \Big[ \frac{\delta E^{\rm dis}}{\delta \rho_{i,\pm}} \Big].
\end{gather}
Analogous to the phase-field energy functional, the energy functional $E^{\rm dis}$ of the dislocation system consists of a local `defect energy' which depends on the dislocation densities, and of the elastic energy: 
\begin{equation}
\label{eq:dislocation_energy_functional}
E^{\rm dis} = \int ({\cal E}^{\rm lda}(\rho_{i,\pm}) + {\cal E}^{\rm el}) \diff V
\end{equation}
where the local density dependent 'defect energy density' has been shown by \citet{Zaiser_2015_PRB} to be of the form
\begin{gather} 
\label{eq:correlation_energy}
{\cal E}^{\rm lda} = \sum_i \left[ Gb^2A\rho_i \ln \big( \frac{\rho}{\rho_0} \big) + \frac{Gb^2D\kappa_i^2}{2\rho} \right],
\end{gather}
where $\rho_i = \rho_{i,+}+\rho_{i,-}$ is the total dislocation density on slip system $i$, $\kappa_i = \rho_{i,+}-\rho_{i,-}$ is the excess density, $\rho = \sum_i \rho_i$ is the overall dislocation density, and $\rho_0$ is a reference density related to dislocation core properties. 

The elastic energy density has exactly the same structure as in the phase field model, in fact, it is the {\em same} physical quantity. Besides the inelastic strains associated with the phase variables, we need now in addition to account for the plastic strain $ \Bepsilon^{\rm pl}$ generated by dislocation motion. This gives for the overall elastic energy density the equation
\begin{equation}
\label{eq:elastic_energy_density2}
 {\cal E}^{\rm el} = \frac{1}{2} (\Bepsilon - \Bepsilon^{\rm inel} - \Bepsilon^{\rm pl}):\IC:(\Bepsilon - \Bepsilon^{\rm inel} - \Bepsilon^{\rm pl}).
\end{equation} 
The plastic strain can be expressed in terms of the shear strains $\eta_i$ on the different slip systems as 
\begin{equation}
\Bepsilon^{\rm pl} = \sum_i {\cal M}_i \eta_i
\end{equation}
where the projection tensors ${\cal M}_i = (\Bg_i \otimes \Bn_i)_{\rm sym}$ are the symmetrized product tensors of the slip plane normal and slip vectors. The shear strain variables $\eta_i$ relate to the excess dislocation densities $\kappa_i = \rho_{i,+}-\rho_{i,-}$ via
\begin{equation}
\kappa_i = - \frac{1}{b} (\Bg_i.\nabla) \eta_i.
\label{eq:gnd}
\end{equation}
Inserting the energy functional, \eqref{eq:dislocation_energy_functional} and \eqref{eq:correlation_energy}, into \eqref{eq:rho_plus_evolution1} and \eqref{eq:rho_minus_evolution1}, and using the relations $\delta/\delta\rho_{i,+} = \delta/\delta \rho + \delta/\delta \kappa$ and 
$\delta/\delta\rho_{i,-} = \delta/\delta \rho - \delta/\delta \kappa$, we obtain the dislocation density transport equations
\begin{eqnarray} 
\label{eq:rho_plus_evolution2}
\partial_t \rho_{i,+} &=& \frac{b}{B} (\Bg.\nabla_{i}) \Big[\rho_{i,+}  \big ( - \tau_i - \tau^{\rm b}_i + \tau^{\rm d}_i \big) \Big], \\
\label{eq:rho_minus_evolution2}
\partial_t \rho_{i,-} &=& \frac{b}{B} (\Bg.\nabla_{i}) \Big[\rho_{i,-}  \big ( \tau_i + \tau^{\rm b}_i + \tau^{\rm d}_i \big) \Big].
\end{eqnarray}
Here, the shear-stress like variables $\tau_i^{\rm b}$ and $\tau_i^{\rm d}$ derive from the defect energy density ${\cal E}^{\rm lda}$ according to
\begin{gather}
\label{eq:stresses}
\tau^{\rm b}_i = GbD \frac{(\Bg_i \cdot \nabla) \kappa_i}{\rho}, \quad \tau^{\rm d}_i =  GbA \frac{(\Bg_i \cdot \nabla) \rho}{\rho}.
\end{gather} 
where we have neglected terms of the order of $(\kappa/\rho)^2$. The resolved shear stresses $\tau_i$ which derive from variation of the elastic energy and using \eqref{eq:gnd} are simply the classical resolved shear stresses in the respective slip systems,
\begin{equation}
\label{eq:rss}
\tau_i = {\cal M}_i:\IC:(\Bepsilon - \Bepsilon^{\rm inel} - \sum_i {\cal M}_i \eta_i).
\end{equation} 
The most convenient method to obtain the plastic shear strains $\eta_i$ is simply to evaluate them on the run while integrating the equations
of evolution of $\rho_{i,\pm}$. By using \eqref{eq:gnd} alongside the definition of $\kappa_i = \rho_{i,+} - \rho_{i,-}$ we get the Orowan-like equation
\begin{equation}
\label{eq:Orowan_equation}
 \partial_t \eta_i  = b \Bg.(\BJ_{\rho_{i,+}} - \BJ_{\rho_{i,-}}).
\end{equation} 
This CDD model is coupled to the phase microstructure in a dual manner: first, the mobility coefficients which control the kinetics of dislocation density evolution evidently depend on the phase microstructure, since both the chemical composition and the structural order of the material strongly influence dislocation mobility. This coupling is unidirectional: The phase microstructure influences the dislocation microstructure evolution but not vice vesa. A bidirectional coupling arises whenever the phase microstructure variables are associated with inelastic strains $\Bepsilon^{\rm inel}$. In this case the elastic energy density, \eqref{eq:elastic_energy_density2}, which is shared by the energy functionals for the dislocation and for the phase microstructure, inevitably contains products of the dislocation density-related strains $\eta_i$ and the phase-variable related strains $\Bepsilon^{\rm inel}$ such that the phase microstructure evolution is influenced by the dislocation configuration and vice versa. Effects of this coupling will now be studied for cyclically loaded Ni based superalloys. 

\subsection{Application of the general phase-field framework to single crystal Nickel-based superalloys}

The Ni-Al binary system of single crystal Nickel-based superalloys mainly consists of cubic $\gp$ phase precipitates surrounded by channel-like $\g$ matrix. For this binary system, only one concentration quantity is need, which is the concentration $c$ of Al atoms. Since $\gp$ phase  has four crystallographic variants, four order parameters $\phi_i, i=1...4$ are used. The phase microstructure is represented as follows: inside $\g$ phase $c$ has the equilibrium concentration $c^{\rm e}_{\g}$ and order parameter $\phi_i=0$; inside $\gp$ phase $c$ has the equilibrium concentration $c^{\rm e}_{\gp}$ and order parameter $\phi_i=1, \phi_j=0, i \neq j$; diffuse interfaces between $\g$ and $\gp$ have $c$ values between $c^{\rm e}_{\g}$ and $c^{\rm e}_{\gp}$, and $\phi_i$ values between 0 and 1. Therefore, the concentration and bulk energy density can be generally constructed in the form of an interpolation between $\g$ and $\gp$ phases, following the spirit of the Kim-Kim-Suzuki Model \citep{Kim_1999_PRE}: 
\begin{equation}
c= \Big( 1-h(\allphi) \Big) \,c_\g + h(\allphi)\,c_\gp ,
\end{equation}
\begin{equation}
\begin{aligned}
{\cal E}^{\rm bulk} = &\Big( 1-h\left(\allphi\right) \Big)\, f^\gamma 
+ h(\allphi)\,f^{\gamma^\prime} \\ 
&+ \omega g(\allphi),
\end{aligned}
\end{equation}
where $h(\allphi)=\sum^{4}_{i=1}[ \phi_i^3(6\phi_i^2-15\phi_i+10) ]$ is a interpolation function which is zero inside $\g$ and one inside $\gp$. $g(\allphi)=\sum^{4}_{i=1} \phi_i^2(1-\phi_i^2) + \theta \sum^{4}_{i,j=1} \phi_i^2 \phi_j^2$ is another interpolation function which penalizes co-existence of different $\phi_i$ and relates to the anti-phase boundary (APB) energy of interfaces between two $\gp$ variants. $f^\g= f_0(c_\g-c_\g^{\rm e})^2$ and $f^\gp = f_0(c_\gp-c_\gp^{\rm e})^2$ are free energies of $\g$ and $\gp$ phase respectively. Local equilibrium requires $\diff f^\g / \diff c_\g = \diff f^\gp / \diff c_\gp$. The gradient energy density, which is constructed as
\begin{equation} 
\label{eq:free_energy}
{\cal E}^{\rm grad} = \frac{k_\phi}{2}\sum^{4}_{i=1}|\nabla\phi_i|^2,
\end{equation} 
controls the $\ggp$ interface properties. The phase associated inelastic strain in the elastic energy density, \eqref{eq:elastic_energy_density2}, is given by
\begin{equation}
\label{eq:inealstic_strain}
\Bepsilon^{{\rm inel}} = h(\allphi) \bar\epsilon^{\rm mis} \matone,
\end{equation}
where $\bar\epsilon^{\rm mis}$ is the eigenstrain due to $\ggp$ lattice misfit and $\matone$ is the second order unit tensor. In evaluating the elastic energy, we use the same stiffness tensor $\IC$ for $\g$ and $\gp$ phases in order to discard effects of elastic inhomogeneity and focus on the role of plasticity. 

Finally, we account for the influence of the phase microstructure on dislocation mobility by setting $B = B_0/[1-h(\allphi)]$ which sets the 
dislocation mobility to zero in the pure $\gp$ phase. This reflects the fact that dislocations entering the $\gp$ precipitates must trail antiphase boundaries which, at the stresses typical of our simulations, renders them effectively immobile inside the precipitates. 

\section{Results and Discussion}
\subsection{Numerics and Simulation setup}
The governing equations of dislocation and phase microstructures are time dependent partial differential equations (PDEs) which contain convection, diffusion and source terms. Given that we use periodic boundary conditions, and both dislocation densities and element concentrations are conserved, Finite Volume Method (FVM) is selected due to its best performance in conservation equations. Nevertheless, FVM is not as powerful as Finite Element Method (FEM) in solving equations in solid mechanics, especially in inclusion problems. The present work therefore uses a numerical scheme coupling FVM and FEM in the same domain: FEM solves the mechanical equilibrium equation,  and FVM solves other equations. We use the same domain (or mesh) in FVM and FEM. The difference is that FVM solves quantity values at element centers, while FEM solves quantity values at vertices (and values at quadrature points are involved in solving process). FVM and FEM are coupled as follows: at each time step, the inelastic strains at the element centers are assigned to quadrature points of the same element for Gaussian quadrature integration, such that the displacements at vertices can be obtained by FEM. We then solve the mechanical equilibrium equations (i.e., we minimize the elastic energy with respect to the displacement field), assuming that mechanical equilibrium is established instantaneously. In implementing periodic boundary conditions for the FEM calculation, an important technical point needs to be taken into account as periodic displacements are not compatible with a periodic strain field of non-zero spatial average: The average of a strain field with periodic displacement BCs is by construction zero. To resolve this problem we need to add the imposed, cyclically alternating average strain `by hand', setting $\Bepsilon \to \Bepsilon + \langle \Bepsilon^{\rm app}(t) \rangle$ and re-evaluating the elastic strain, and thus the stress, accordingly. The displacements strains and stresses calculated in this manner are interpolated back to the element centers by using the FEM shape functions, and thereby transferred to the FVM algorithm for solving the microstructure evolution equations in the next time step. A detailed numerical procedure can be found in \cite{Wu_2017_PhDthesis}. 

In our simulations we consider a single crystal of a NiAl superalloy with $\g\gp$ microstructure that is cyclically loaded in [010] lattice direction of the fcc crystal lattice. We assume plane strain conditions, imposing homogeneity of both the dislocation and the phase microstructure in [001] direction. This is, of course, a strong assumption since real $\g\gp$ microstructures are three-dimensional, however, it allows us to map the dynamics of dislocations onto a conserved dynamics of point-like objects in an intersecting $(001)$ plane and thus use the formalism described in the previous section. The assumption of plane strain deformation implies that from eight potentially active slip systems in a [100] oriented crystal, only four contribute to the deformation. These are the slip system pair $[110](1\bar{1}1)$ with shear strain $\eta_{\rm I}$ and $[110](1\bar{1}\bar{1})$  with shear strain $\eta_{\rm II}$ which are symmetrically active such that $\eta_{\rm I} = \eta_{\rm II} = \eta_1/2$, and the slip system pair $[1\bar{1}0](111)$ with shear strain $\eta_{\rm III}$ and $[1\bar{1}0](11\bar{1})$  with shear strain $\eta_{\rm IV}$, which are again symmetrically active such that $\eta_{\rm III} = \eta_{\rm IV} = \eta_2/2$. The overall plastic strain tensor has then the structure
\begin{equation}
\Bepsilon^{\rm p} = \frac{\eta_1 + \eta_2}{6} \left[\begin{array}{ccc}1 & 0  & 0\\ 0 & -1 & 0 \\ 0 & 0 & 0 \end{array} \right],
\end{equation}
showing the plane strain nature of the deformation state.

In a (001) plane of intersection, the intersection points of dislocations of the first pair of active slip systems move along the [110] direction. We combine these slip systems into a single effective slip system $[110](1\bar{1}0)$. Dislocations which thread the plane of intersection in the positive [001] direction are defined as positive edge dislocations of the effective slip system, and dislocations which thread the plane of intersection in the negative [001] direction are defined as negative edge dislocations (note that the actual dislocations possess screw parts which are, however, of equal density and opposite sign such that they cancel from the stress and strain calculations). Similarly, we combine the second pair of active slip systems into an effective slip system $[1\bar{1}0](110)$. In the following we  drop the irrelevant $x_3$ coordinate when labeling lattice directions.

This situation is illustrated in  \figref{fig:fatigue_simulation_schematic} left where we have dropped the irrelevant $x_3$ coordinate. We simulate a representative mesoscopic area in a macroscopic sample, so periodic boundary conditions are used for the system. The external loading direction is aligned with the [01] crystal orientation. $\gp$ precipitates are surrounded by channel-like $\g$ phase. The $\ggp$ interfaces are labeled with "up", "right", "down" and "left" for distinguishing purposes. Densities of dislocations moving in [11] and [$\bar{1}$1] directions are distinguished by the subscript $k \in \{1,2\}$, the respective dislocation densities $\rho_{k,+}=\rho_{k,-}=\SI{1.25e13}{m^{-2}}$ are chosen such that the total dislocation density agrees with typical values observed in experiments \cite{Yoshihiro_2014_proceeding, Jacome_2013_AM}. Since there are equal densities of positive and negative dislocations and dislocations are initially homogeneously distributed over the slip channels, the initial excess dislocation density $\kappa_{k}$ and shear strain $\eta_{k}$ are both zero. Other simulation parameters are listed in \tabref{tab:Fatigue_Simulation_Parameters}. 

The assumption of plane strain which allows us to represent the dislocation state in terms of a 2D system of dislocations pertaining to two effective slip systems is, of course, not fully realistic. As a consequence, all comparisons with experimental data should be considered in qualitative, rather than quantitaitve terms. Perhaps the most significant difference between 2D and 3D dislocation systems concerns effects
of curvature, since the dislocation intersections with a representative plane actually form parts of three-dimensionally curved loops \cite{Jacome_2013_AM}. The motion of such loops -- and therefore the flow stress of the $\g$ channels -- is controlled by Orowan stresses, and indeed enhanced Orowan strengthening has been a major driver for the reduction in scale of $\ggp$ microstructures over the past decades of superalloy development. In the present 2D model, line tension cannot be directly represented. However, the back stress terms $\tau_{\rm b}$ which derive, in the present work, from the defect energy functional lead to a very similar behavior:  \citet{Forest_2003_PM} compared models containing back stress terms, which are according to \eqref{eq:stresses} proportional to the gradient of $\kappa$ and hence because of \eqref{eq:gnd} to the second gradient of the plastic shear strain, with models that explicitly account for line tension of curved dislocations. They showed that both types of models yield similar predictions regarding plastic flow in confined channels, which is the main focus of interest in plasticity of $\ggp$ superalloys. Hence, we can indirectly account for Orowan effects by adjusting the coupling parameter $D$, \eqref{eq:stresses}, in such a manner as to correctly reproduce the channel size dependent flow behavior of experimental $\ggp$ microstructures.   

\begin{table}
	\centering
    \small
	\captionof{table}{Simulation parameters \citep{Zhou_2007_AM, Zaiser_2014_MSMSE, Sills_2016_PM}} 
	\label{tab:Fatigue_Simulation_Parameters}
\begin{tabular}{l@{\hskip 0.02in}l}
	\hline
	{equilibrium concentration of $\g$ }  & $c^{\rm e}_{\g}$  = $0.160$  \\
	{equilibrium concentration of $\gp$}  & $c^{\rm e}_{\gp}$ = $0.229$  \\
	{bulk energy coefficient}             & $f_0$       = $3.2\cdot10^{9}$   \, \SI{}{J m^{-3}} \\
	gradient energy coefficient           & $k_\phi$       = $9.4\cdot10^{10}$  \,\SI{}{J m^{-1}} \\
	interface energy coefficient          & $\omega$    = $3.9\cdot10^{6}$   \,\SI{}{J m^{-3}} \\
	APB energy coefficient                & $\theta$    = $10$               \, \\
	{diffusion mobility coefficient}      & $M_c$         = $1.5\cdot10^{-26}$ \SI{}{m^{5}J^{-1}s^{-1}}\\
	{ordering mobility coefficient}       & $M_\phi$         = $5.8\cdot10^{-9}$  \SI{}{m^{3}J^{-1}s^{-1}}\\
	{component of stiffness tensor}       & $C_{11}$    = $163$              \,\SI{}{GPa}\\
	{component of stiffness tensor}       & $C_{12}$    = $112$              \,\SI{}{GPa}\\
	{component of stiffness tensor}       & $C_{44}$    = $86$               \,\SI{}{GPa}\\
	{$\ggp$ misfit strain}                & $\bar\epsilon^{\rm{mis}}$ = $-0.003$ \,      \\
	{magnitude of Burgers vector}         & $b$         = $0.25$             \,\SI{}{nm} \\
	{back stress coefficient}             & $D$         = $0.6$              \,  \\
	{diffusion stress coefficient}        & $A$         = $0.6$              \,  \\
	{yield stress coefficient}            & $\alpha$    = $0.3$              \,  \\
	{drag coefficient}                    & $B_0$         = $4\cdot10^{-6}$ \,\SI{}{GPa\,s} \\ \hline
\end{tabular}
\end{table}

To reveal the loading parameter-microstructure-mechanical property relations, six simulations are designed as in \tabref{tab:Fatigue_Simulation_Design}. "S0" is the case of no applied stress and strain, the microstructure of "S0" serves as a reference. Four significant cyclic loading parameters are considered in the present work: Ratio (of compressive to tensile peak strain), strain range, waveform and period. While studying the role of any one of these, the remaining parameters are kept fixed. Therefore, comparison of "S1" and "S2" gives the effect of ratio, "S1" and "S3" gives the effect of range, "S1" and "S4" gives the effect of waveform, "S1" and "S5" gives the effect of period. 

\begin{table}
	\centering
    \small
	\captionof{table}{Simulation design} 
    \label{tab:Fatigue_Simulation_Design}
	\begin{tabular}{ccccc}
		\hline
		        & Ratio & Range (\%) & Waveform & Period (s) \\
		\hline
		S0                & -  & -   & -           & -  \\
		S1                & 0  & 0.4 & Triangular  & 40  \\
		S2                & -1 & 0.4 & Triangular  & 40  \\
		S3                & 0  & 0.2 & Triangular  & 40  \\
		S4                & 0  & 0.4 & Trapezoidal & 40  \\
		S5                & 0  & 0.4 & Triangular & 400  \\
		\hline
	\end{tabular}
\end{table}
\subsection{Short term microstructure-property relations within one cycle}
\begin{figure}[htp] \centering
	\includegraphics[width=\columnwidth]{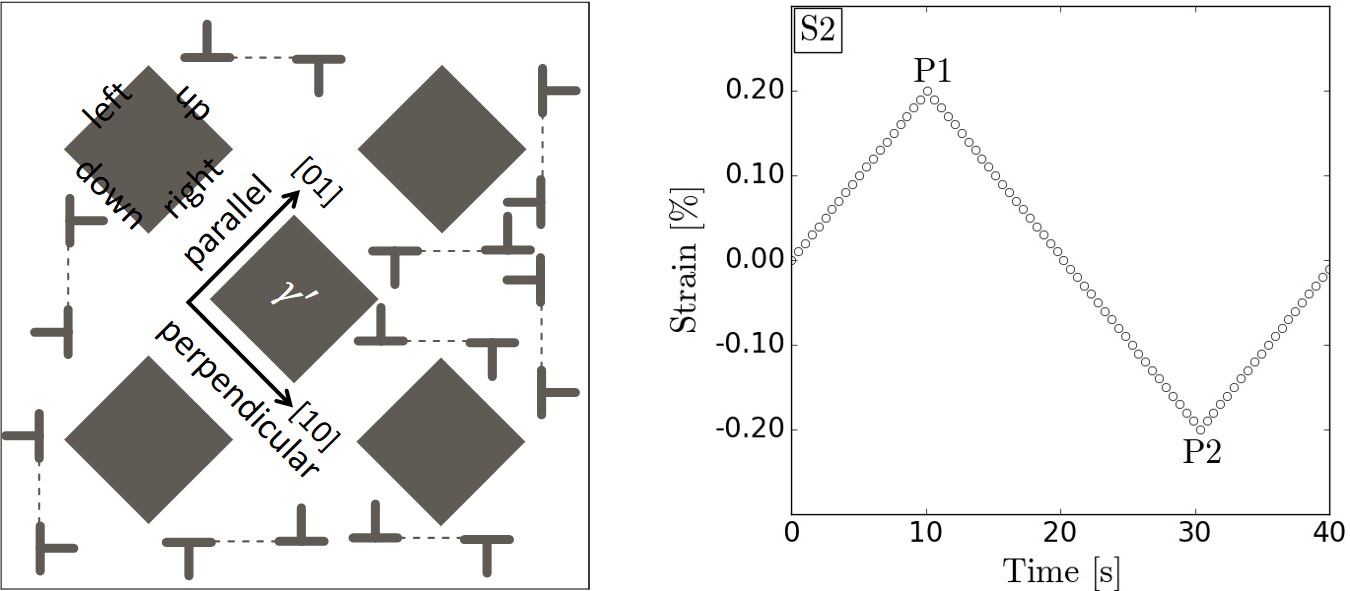}
	\caption{\label{fig:fatigue_simulation_schematic} Left, schematic of simulated system; right, waveform of simulation "S2".}
\end{figure}

\begin{figure}[htp] \centering
	\includegraphics[width=\columnwidth]{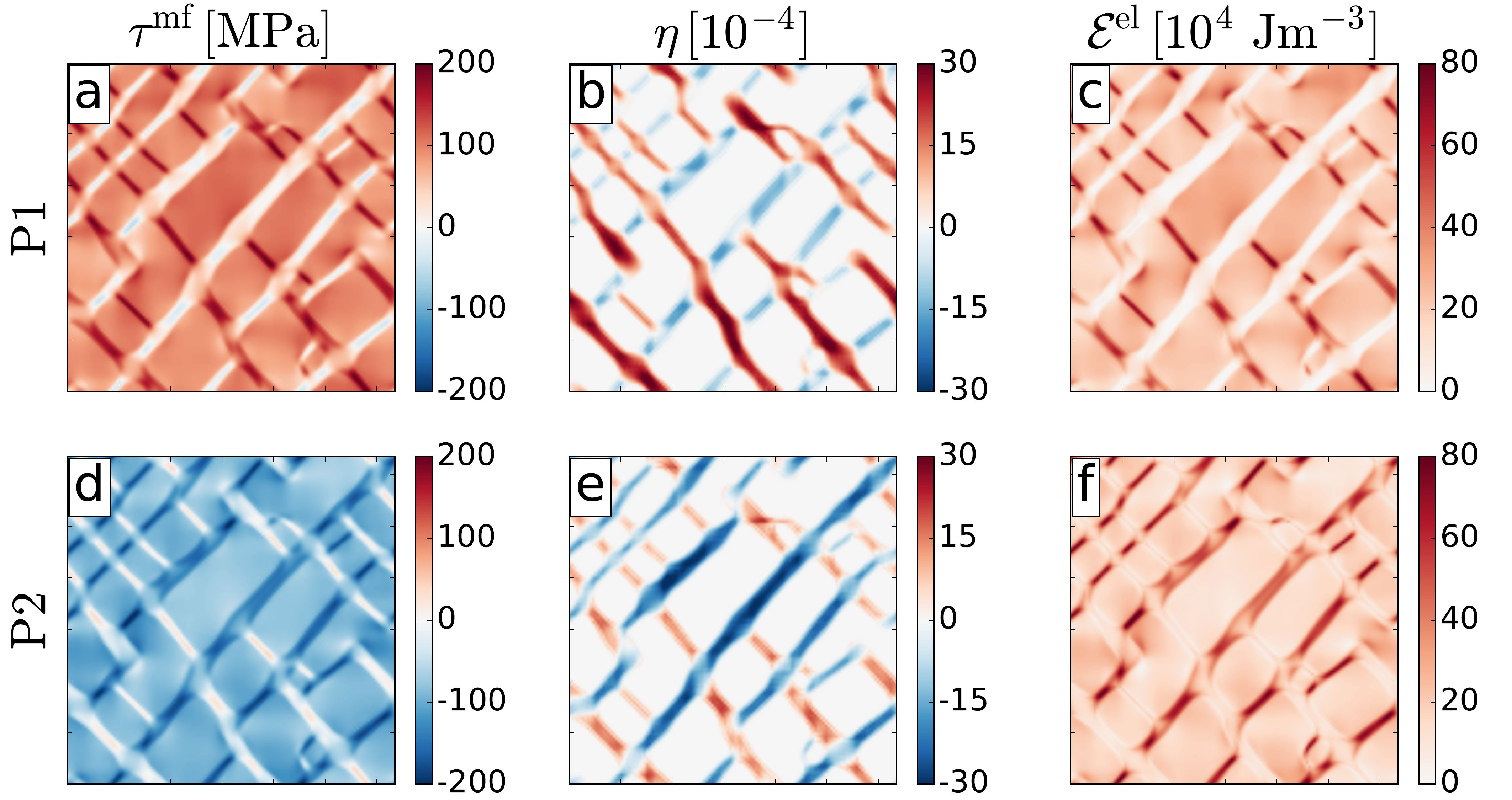}
	\caption{\label{fig:fatigue_dislocation_field} Spatial distribution of some field variables as obtained from simulation "S2" during the first deformation cycle: (a) resolved shear stress $\tau$ at loading peak "P1", note that $\tau_1 = \tau_2$; (b) shear strain $\eta = \eta_1 + \eta_2$ at "P1", (c) elastic energy density $\cal E^{\rm el}$ at "P1", (d) $\tau$ at loading valley "P2", (e) $\eta$ at "P2", (f) $\cal E^{\rm el}$ at "P2". }
\end{figure}

We take "S2" as an example for explaining the precipitate-dislocation interaction mechanism. The waveform of "S2" is plotted in \figref{fig:fatigue_simulation_schematic} right, and some quantities at peak loading point "P1" and valley point "P2" are shown in \figref{fig:fatigue_dislocation_field}. At loading peak "P1", the perpendicular $\g$ channels exhibit a large (about \SI{200}{MPa}) positive resolved shear stress, while parallel $\g$ channels exhibit a small negative resolved shear stress (see \figref{fig:fatigue_dislocation_field} (a)). This results in positive shear in perpendicular $\g$ channels and negative shear in parallel $\g$ channels, and the magnitude of the positive shear is much bigger than of the negative (see \figref{fig:fatigue_dislocation_field} (b)). The positive shear in perpendicular $\g$ channels implies that positive dislocations move to down-perpendicular $\ggp$ interfaces while negative dislocations move to up-perpendicular $\ggp$ interfaces. Since the dislocation mobility is zero inside $\gp$ precipitates, dislocations eventually pileup at $\ggp$ interfaces. In parallel $\g$ channels, negative shear implies positive dislocations pileup at right-parallel $\ggp$ interfaces while negative dislocations pileup at left-parallel $\ggp$ interfaces. As the positive shear in perpendicular $\g$ channels is much more pronounced than the negative shear in parallel $\g$ channels, the dislocation pileup is accordingly more pronounced in perpendicular $\g$ channels. The dislocation pileup structures at $\ggp$ interfaces alter the elastic energy density, which in perpendicular $\g$ channels is much higher than in parallel $\g$ channels (see \figref{fig:fatigue_dislocation_field} (c)). Since dislocation activities do not change bulk and gradient energy of phase microstructures, the alteration of elastic energy plays a major role in determining coarsening behaviour of $\gp$ precipitates. \citet{Cottura_2015_AM} has demonstrated that the minimum elastic energy is matter of distance between $\ggp$ precipitates: too close and too far $\gp$ distances both increase elastic energy. Therefore, $\gp$ precipitates would evolve in a way to adjust their distances such that the minimum elastic energy is approached. $\gp$ precipitates adjust their distances by diffusion-controlled directional motion of $\ggp$ interfaces, which also appears as rafting (directional coarsening) of $\gp$ precipitates. Specifically, for the elastic energy density in \figref{fig:fatigue_dislocation_field} (c), $\gp$ precipitates tend to raft to perpendicular direction such that the perpendicular channels are widened and elastic energy density is relaxed. Even though simultaneously, parallel channels are narrowed and elastic energy density there is increased, the total elastic energy is decreased, as has been demonstrated by \citet{Wu_2017_IJP}. In contrast, at loading valley "P2", the characteristics (see \figref{fig:fatigue_dislocation_field} (d)-(f)) are basically opposite to those observed at the point "P1". 

\begin{figure}[htp] \centering
	\includegraphics[width=\columnwidth]{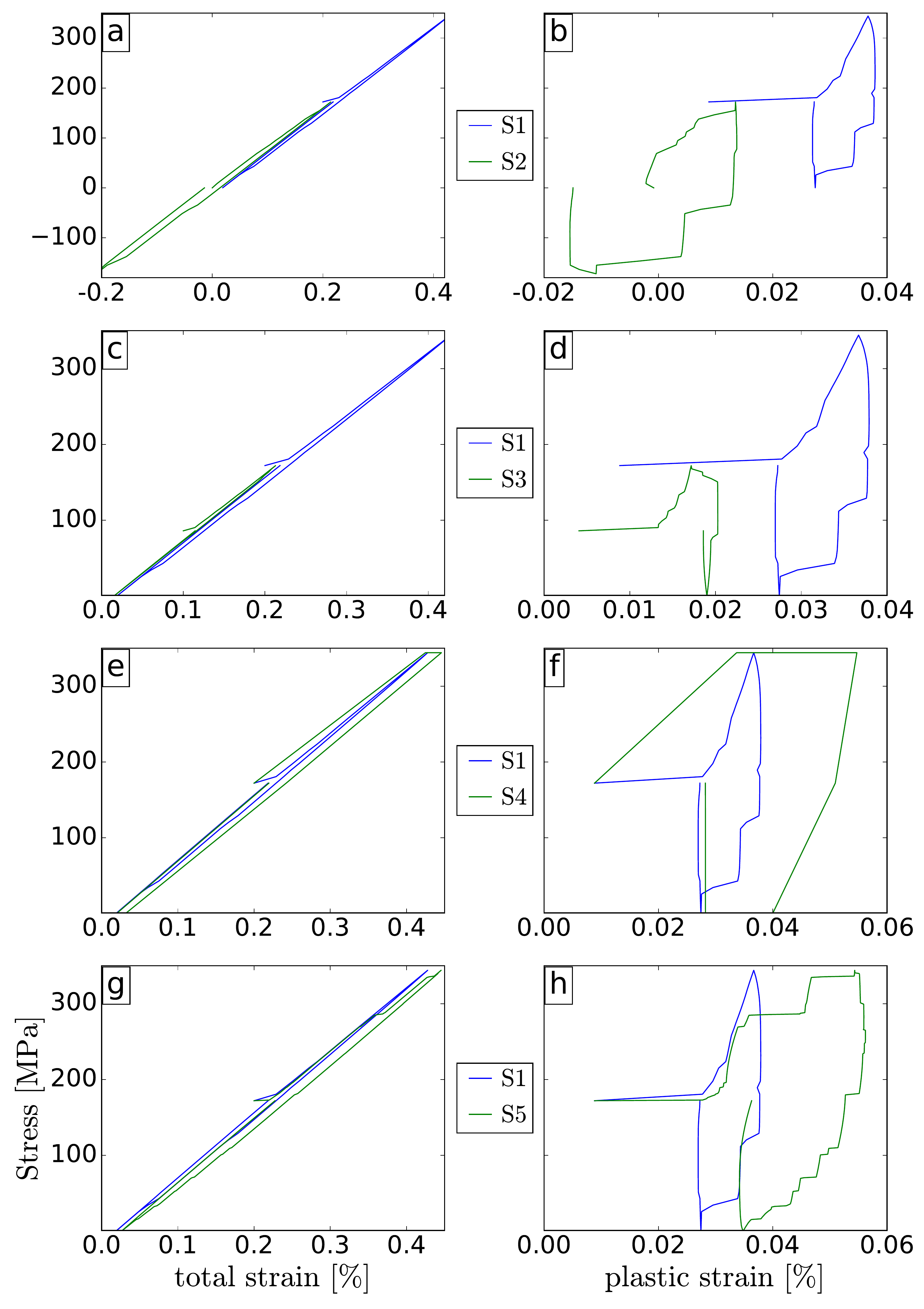}
	\caption{\label{fig:fatigue_loop_of_first_cycle} Cyclic loops at first cycle: (a) total strain of "S1" and "S2", (b) plastic strain of "S1" and "S2", (c) total strain of "S1" and "S3", (d) plastic strain of "S1" and "S3", (e) total strain of "S1" and "S4", (f) plastic strain of "S1" and "S4", (g) total strain of "S1" and "S5", (h) plastic strain of "S1" and "S5". }
\end{figure}

Without any additional assumptions, we obtain cyclic stress-strain loops as natural results of precipitate-dislocation interactions and the corresponding evolution of the internal stress and strain fields under cyclic loading, as shown in \figref{fig:fatigue_loop_of_first_cycle} for the first cycle loops of all simulations. One common feature in \figref{fig:fatigue_loop_of_first_cycle} is that total strains at the same stress under loading and unloading are close, which leads to nearly linear relations between stress and total strain. The reason is that during one cycle, the phase microstructure and therefore $\ggp$ misfit strain alter negligibly due to the diffusion-controlled nature of phase microstructure evolution, meaning that the change in strain mainly comes from the change in plastic strain due to back-and-forth dislocation motion. Comparison of the total strain (left column of \figref{fig:fatigue_loop_of_first_cycle}) and the plastic strain (right column of \figref{fig:fatigue_loop_of_first_cycle}) shows that plastic strain is one order of magnitude smaller than total strain. By comparing "S1" with other simulations, the influence of different loading parameters can be studied. Shifting cyclic ratio from $0$ to $-1$ leads, as it should be, to a shift of mean strain and stress (see \figref{fig:fatigue_loop_of_first_cycle} (a) and (b)). Decreasing the strain range from $0.4$ to $0.2$, apart from decreasing the stress range, reduces  the strain difference between loading and unloading at the same stress (see \figref{fig:fatigue_loop_of_first_cycle} (c) and (d)). In contrast, an increasing strain difference between loading and unloading at the same stress can be seen when changing the waveform from triangular to trapezoidal (see \figref{fig:fatigue_loop_of_first_cycle} (e) and (f)). A similar feature is also present if increasing the period from \SI{40}{s} to \SI{400}{s} (\figref{fig:fatigue_loop_of_first_cycle} (g) and (h)). Moreover, an interesting phenomenon is that the intermittent step-like avalanche is more pronounced in "S5" (see \figref{fig:fatigue_loop_of_first_cycle} (h)), because of much smaller strain increment between two time steps in "S5".

\subsection{Long term microstructure and property evolution}

\begin{figure}[htp] \centering
	\includegraphics[width=\columnwidth]{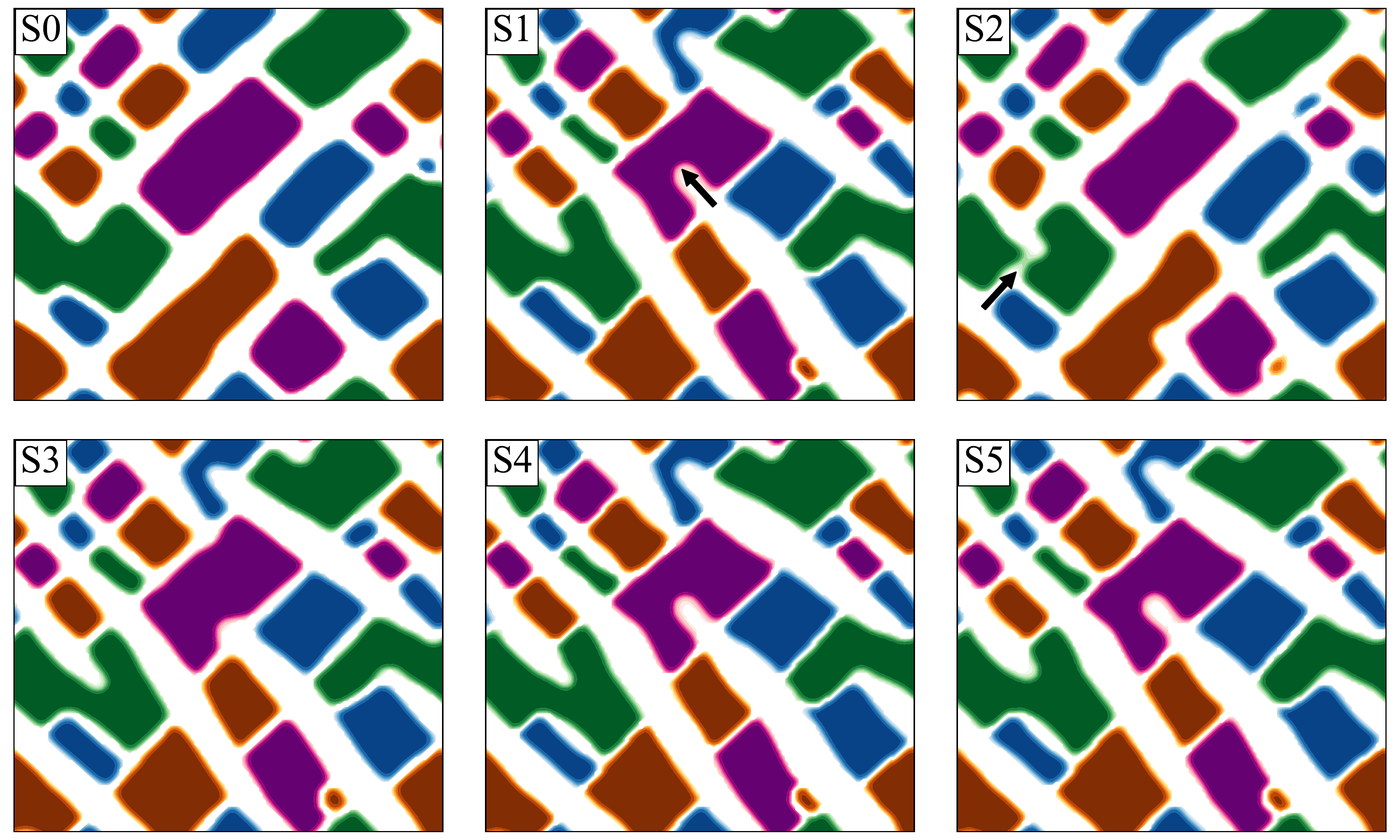}
	\caption{\label{fig:fatigue_phase_field} $\ggp$ microstructures after \SI{20000}{s}, the white background indicates the $\g$ phase, and the four colors indicate four variants of $\gp$ phase.}
\end{figure}

So far we have illustrated the microstructure interactions and resulting mechanical properties in the first cycle. However, more important for the performance of superalloys is the long term co-evolution of microstructure and deformation properties. Different cyclic loadings may result in different microstructures and properties. The $\ggp$ microstructures after \SI{20000}{s} deformation are shown in \figref{fig:fatigue_phase_field}. In comparison with initial microstructures, although $\gp$ precipitates are all coarsened to some extent because of the general tendency to reduce $\ggp$ interface area and therefore total energy, the coarsening ways are different. $\ggp$ of "S0", which acts as a reference of non-loaded microstructure, shows no preferred coarsening direction. One may find several parallel elongated $\g$ precipitates in "S0", however, this feature may occur with equal probability in either [01] or [10] direction and is therefore not a sign of directional coarsening (rafting) \citep{Cottura_2015_AM}. In contrast, in "S1" coarsening in perpendicular, i.e. [10] direction can be seen. This is because the mean loading stress of "S1" is positive and the overall influence of the external loading on the microstucture is similar to creep. The perpendicular and parallel channels are dramatically widened and narrowed, respectively. Another feature is that originally parallel elongated $\gp$ precipitates tend to split, as indicated by the arrow in "S1". Moreover, the $\ggp$ in "S0" are well aligned with perpendicular and parallel directions, while the perpendicular $\ggp$ interfaces and $\g$ channels become more wavy in "S1". The mean stress in "S2" is zero, therefore the "S2" microstructures are similar to "S0", with the minor difference that "S2" has a slower coarsening kinetics as can be seen from the merging precipitates indicated by arrows. However, the comparison of "S1" and "S2" shows that the cyclic ratio has dramatic influence on $\ggp$ microstrucutres. Simulations "S3", "S4" and "S5" show similar features with "S1", implying that $\ggp$ microstrucutres are not very sensitive to cyclic range, waveform and period. Their small differences lie on the extend of rafting: exposure at higher stress or for longer time results in more rafting. 

\begin{figure}[htp] \centering
	\includegraphics[width=\columnwidth]{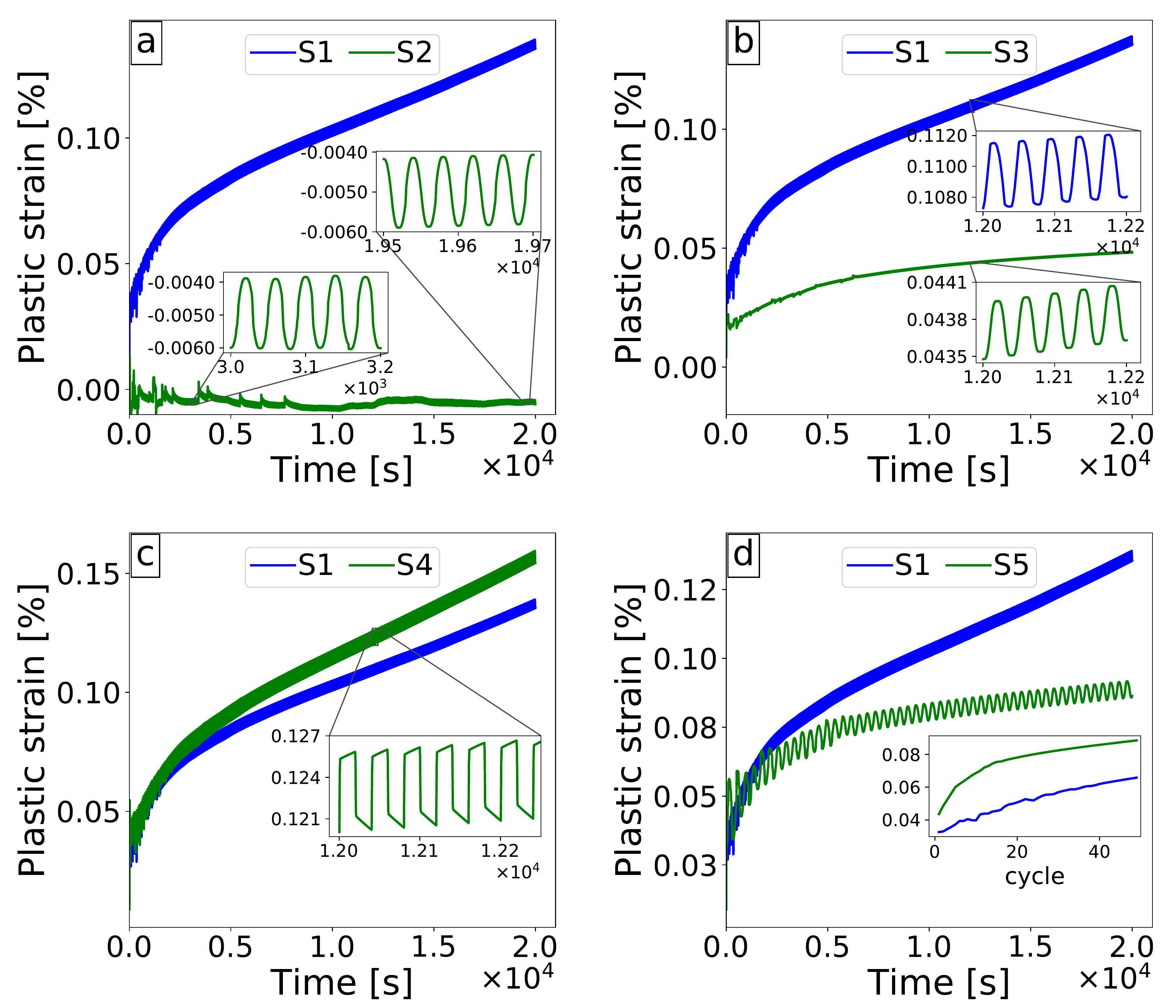}
	\caption{\label{fig:fatigue_curve_VS_time} Evolution of plastic strain with time: (a) "S1" and "S2", insets: magnifications for S2 at two different cycle numbers; (b) "S1" and "S3", insets: magnifications for S1 and S3 at the same time; (c) "S1" and "S4", insets: magnifications for S1 and S4 at the same time;  (d) "S1" and "S5", inset: mean cyclic strain vs. cycle number.}
\end{figure}

Different cyclic loadings may lead to different long term properties. The evolutions of plastic strain with time are plotted in \figref{fig:fatigue_curve_VS_time}. The plastic strain of "S1" shows initially a rapid increase due to massive dislocation motion in $\g$ channels, followed by a steady and slow increment due to the motion of the $\ggp$ interfaces and widening of the $\g$ channels. This trend is similar to the plastic strain evolution in creep deformation, while one evident difference is that plastic strain actually increases in wavy manner due the back-and-forth dislocation motion under cyclic loading. Since the ratio of "S2" is $-1$, the plastic strain basically fluctuates around zero, \figref{fig:fatigue_curve_VS_time} 9(a). As shown in the inset of that figure, also the amplitude of the oscillation is not increasing during prolonged cycling, indicating essentially stationary behavior. The plastic strain tendency of "S3", no matter in long time or short time perspective, resembles "S1" (see \figref{fig:fatigue_curve_VS_time}(b)). However, both the plastic strain magnitude and wave amplitude of "S3" are much smaller than in "S1". On the contrary, the differences between "S1" and "S4" are within $20\%$ in terms of both plastic strain magnitude and wave amplitude (see \figref{fig:fatigue_curve_VS_time}(c)). Whereas their short time perspectives (the zoom-in) show different wave shapes which are analogous to each of their waveforms. The slightly higher plastic strain of "S4" is due to longer exposure time at peak stress of trapezoidal compared with triangular waveform. Comparison of "S1" and "S5" shows that their plastic strains basically overlap in the fast increment stage, while "S1" has considerably higher rate in the following steady increment stage (see \figref{fig:fatigue_curve_VS_time}(d)). The plastic strain evolution with cycle number (cycle-averaged plastic strain vs cycle number) is also plotted. Since the period of "S5" is ten times that of "S1", at the same cycle number the corresponding time of "S5" is also ten times of "S1", leading to higher plastic strain for "S5".       

\begin{figure}[htp] \centering
	\includegraphics[width=\columnwidth]{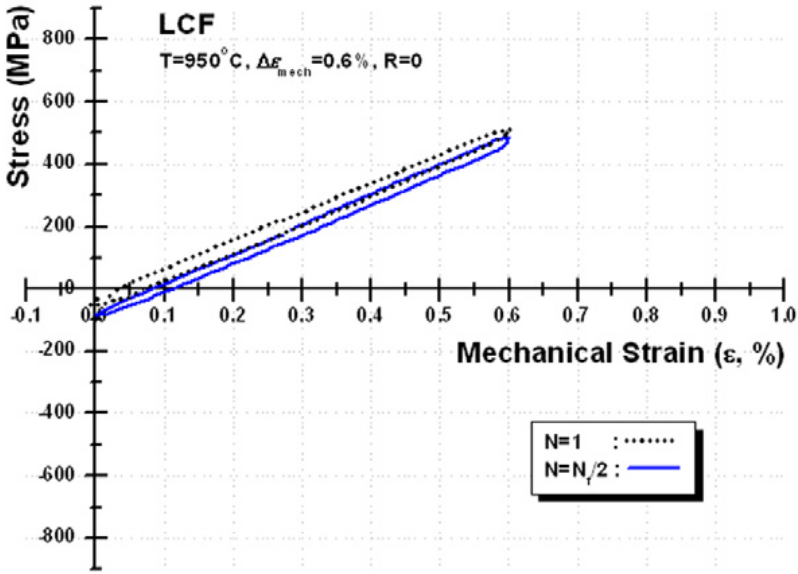}
	\caption{\label{fig:experimental_cyclic_curve} Typical hysteresis loops of isothermal cyclic loading of single crystal Nickel-based superalloys with loading condition: ratio is 0, range is 0.6\%, period is 4 s, waveform is trapezoidal \cite{Hong_2011_IJF}.}
\end{figure}

There exists a significant body of experimental work on cyclic loading of single crystal Nickel-based superalloys, however, the assumption of plane strain deformation made in our work allows only for qualitative or semi-quantitative comparison between experiments and the present simulations. In terms of microstructure, \cite{Hong_2011_IJF} shows that when the tension/compression ratio is 0, significant rafts perpendicular to the loading axis are developed whereas such rafts are absent when the tension/compression ratio is -1. Our simulated microstructures show the same feature, as can bee seen in \figref{fig:fatigue_phase_field}. In terms of mechanical properties, almost all the experimental investigations provide total strain only and do not provide plastic strain. Hence, we focus on the comparison of stress VS total strain between our simulations and \cite{Hong_2011_IJF}. Typical hysteresis loops of isothermal cyclic loading of single crystal Nickel-based superalloys with loading condition similar to "S4" are shown in \figref{fig:experimental_cyclic_curve}, where two features can be seen: i) the loading curve and unloading curves are very close to each other; ii) the curves are nearly linear with a slope about stress/total strain=200 MPa/0.3\%. The same features can be seen in the hysteresis loop of "S4" in \figref{fig:fatigue_loop_of_first_cycle} (e). Therefore, we think it is fair to say that our simulations agree with experiments well, both in terms of microstructure and in terms of mechanical behavior.

\section{Conclusion}
A mesoscale phase-field type model for cyclic loading has been developed. Applying the developed model to high temperature cyclic loading of single crystal Nickel-based superalloys, the simulated results show that the developed model can capture the interactions between of $\ggp$ phase and dislocation microstrucutres, and obtain mechanical properties as the natural outcome of microstructure evolution. Thus, the cyclic loading-microstructure-property relations could be revealed, which are summarised as following:\\
1) In the short term perspective (in one cycle), dislocations move back and forth, leading to cyclic loops in consistent with characteristics observed in experiments. The plastic strains are one order of magnitude smaller than total strains, which explains why the cyclic loops are very "thin".\\
2) In the long term perspective, all $\ggp$ microstructures exhibit directional coarsening similar to creep under zero cyclic ratio, with the extent of rafting slightly depending on waveform, period, etc. The plastic stains are sensitive to cyclic loading conditions both in terms of curve shape and in terms of magnitude. \\
3) Semi-quantitative comparison between experiments and the present simulations shows that simulations results agree with experiments well, both in terms of microstructure and in terms of mechanical properties. 

\section*{Acknowledgment}
Financial support from the following three projects is gratefuly acknowledged: 1) FOR1650 'Dislocation-based Plasticity' (DFG grant ZA171/7-1); 2) the State Key Laboratory of Solidification Processing in Northwestern Polytechnical University, project NO. SKLSP201833; 3) the Fundamental Research Funds for the Central Universities in Northwestern Polytechnical University, project NO. 31020180QD088.

\section*{References}
\bibliography{references}

\end{document}